\documentstyle[twoside,fleqn,espcrc2,epsf]{article}

\newcommand{\AmS}{{\protect\the\textfont2
  A\kern-.1667em\lower.5ex\hbox{M}\kern-.125emS}}

\title{ Visualization of topological objects in the deconfinement phase 
         of pure QCD
\thanks{Supported in part by FWF under Contract No. P11456
        Talk presented by H. Markum}
} 

\author{Markus Feurstein, Harald Markum and Stefan Thurner \\
\vspace{3mm}
Institut f\"{u}r Kernphysik, TU Wien,
         Wiedner Hauptstra\ss e 8-10, A-1040 Vienna, Austria\\
}       
\begin{document}

\begin{abstract}
In the last years it turned out that instantons and monopoles have a 
certain local correlation in
four-dimensional QCD. It was demonstrated by several groups 
independently and by different
methods that at the locations of instantons also monopoles in 
the maximal abelian projection can
be found with enhanced probability. Further we observed such nontrivial 
correlation functions in the
deconfinement phase. We continue our visualization project to 
analyze several specific gauge-field configurations. 
\end{abstract}

\maketitle

There are two different types of topological objects 
which seem to be important candidates for the confinement mechanism:
color magnetic monopoles and instantons.
In lattice calculations we demonstrated that color magnetic monopoles
and instantons are correlated on realistic gauge-field configurations 
\cite{wir}. 
\begin{figure*}[t]
\begin{center}
\begin{tabular}{cc}
\hspace{0.3cm}  Confinement &
\hspace{0.3cm}  Deconfinement\vspace{0.3cm}\\
\epsfxsize=5.5cm\epsffile{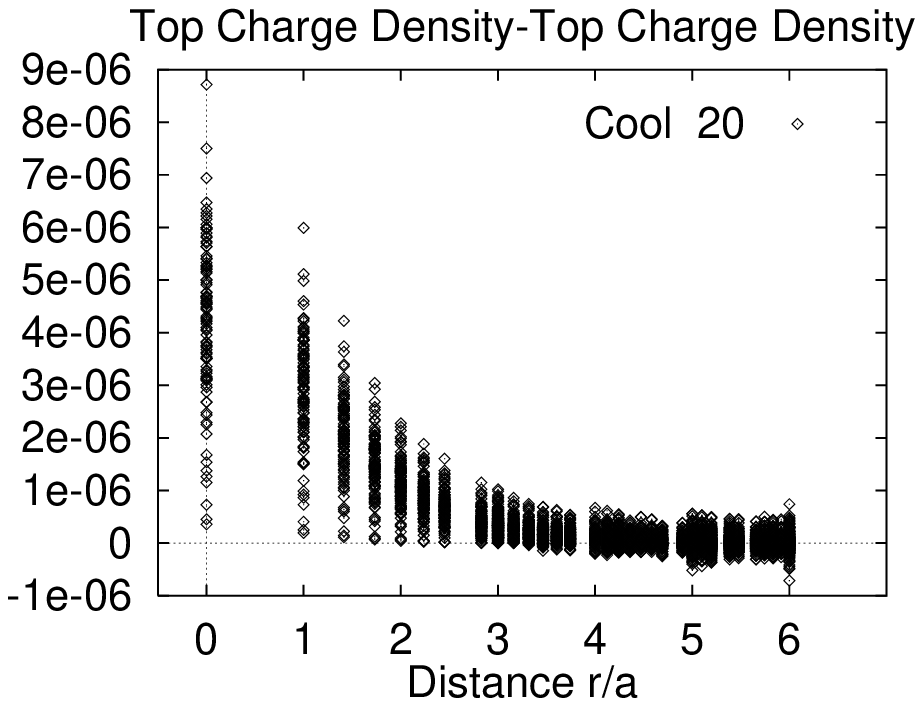} &
\epsfxsize=5.5cm\epsffile{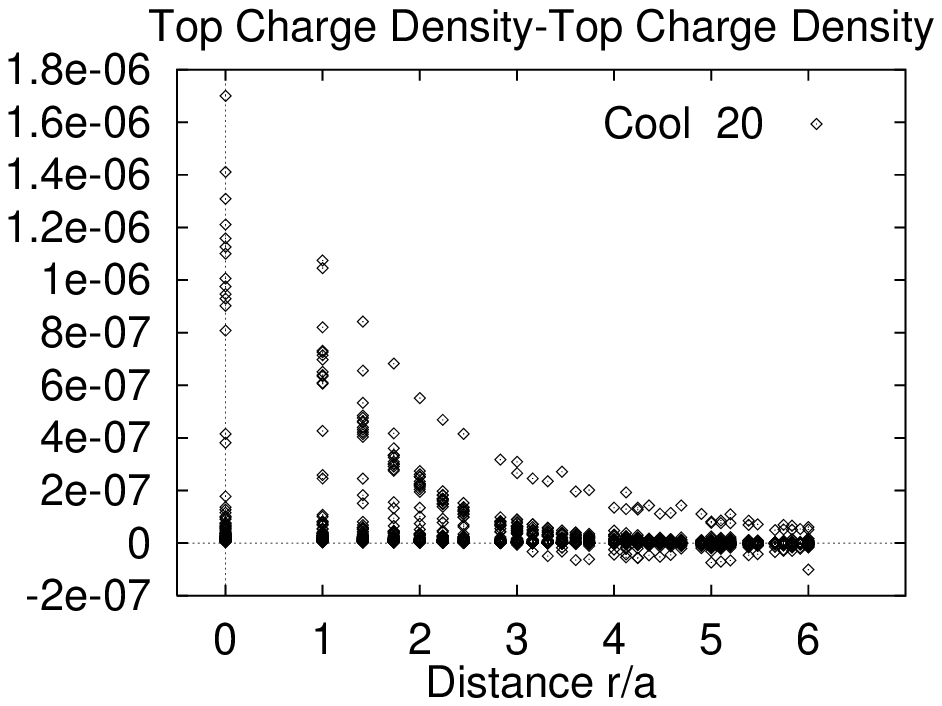} \\  
\epsfxsize=5.5cm\epsffile{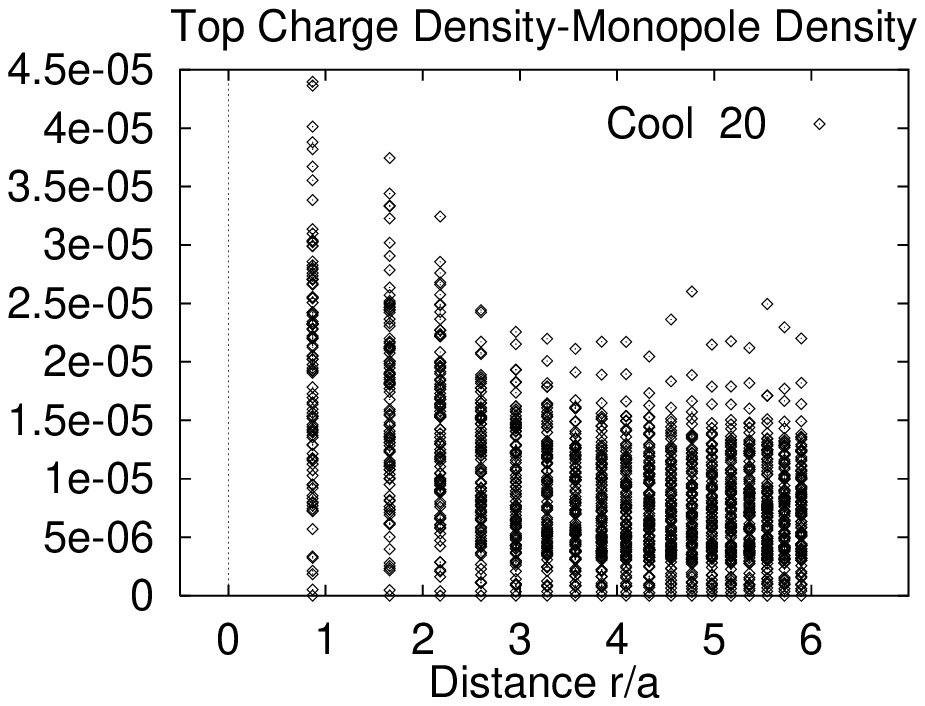} &
\epsfxsize=5.5cm\epsffile{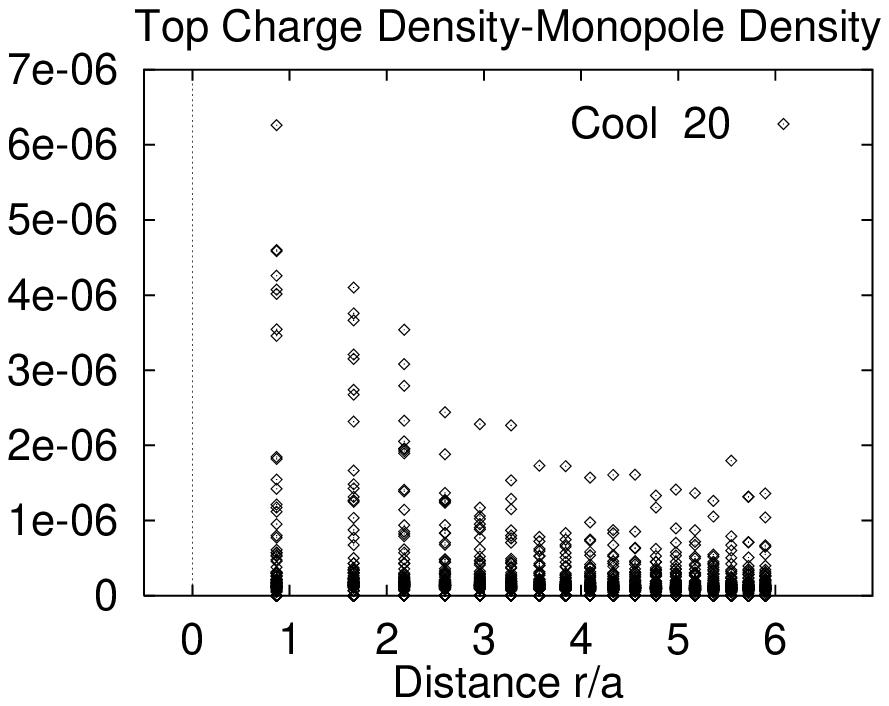} \\ 
\end{tabular}
\end{center}
\vspace{-1.0cm}
\caption{
Auto-correlation functions of the topological charge density
after 20 cooling steps for 100 independent configurations in both phases (top).
In contrast to the confinement phase only 15 \% of the
configurations carry a topological charge in the deconfinement phase.
The corresponding $\rho|q|$-correlations are displayed in both phases (bottom).
All configurations
with  nonvanishing $qq$-auto-correlation
give rise to a nontrivial $\rho|q|$-correlation.
} 
\end{figure*}
\begin{figure}[ht!]
\begin{center}
\begin{tabular}{c}
\bf    t = 2 fixed \\
\epsfxsize=5cm\epsffile{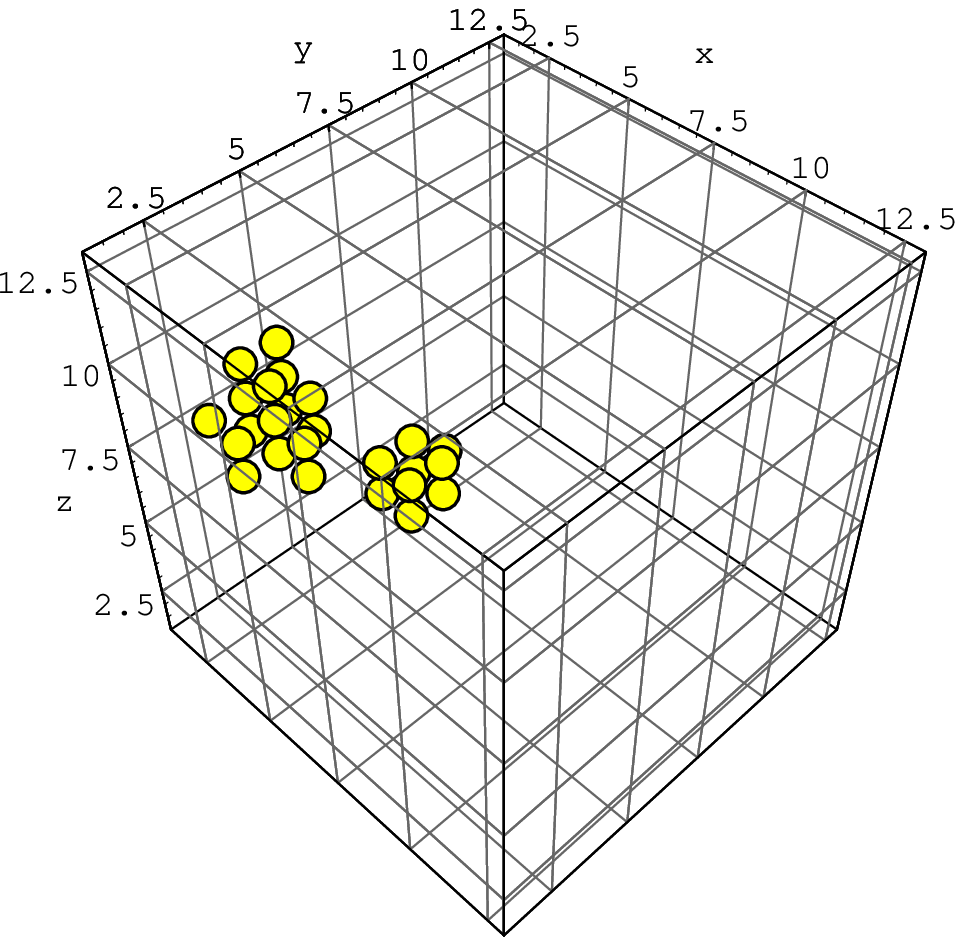} \vspace{0.0cm} \\
\bf    x = 3 fixed \\
\epsfxsize=5cm\epsffile{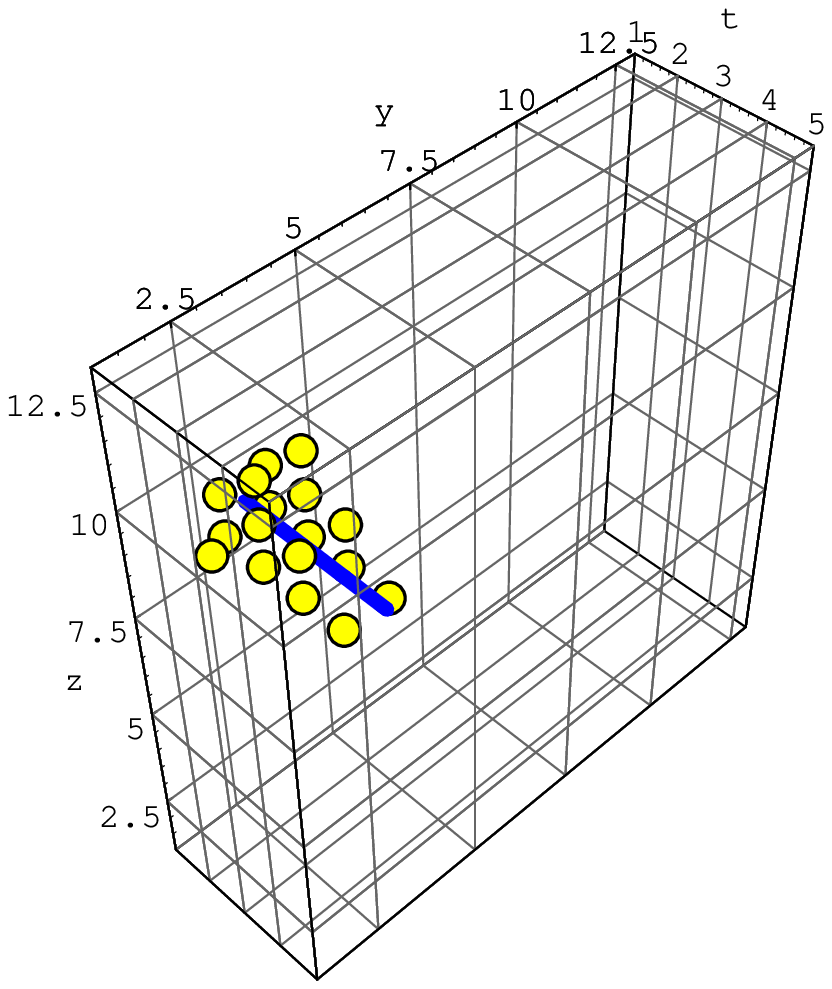} \vspace{0.0cm} \\ 
\bf    x = 5 fixed \\
\epsfxsize=5cm\epsffile{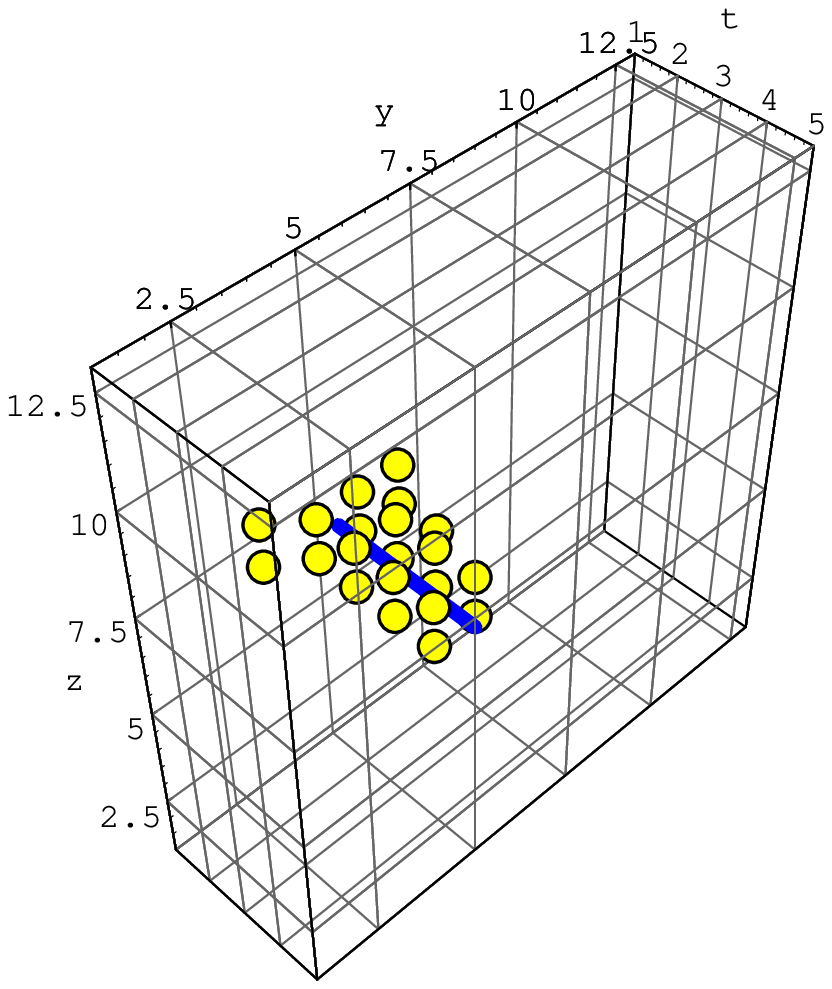} \vspace{0.0cm} \\ 
\end{tabular}
\end{center}
\vspace{-1.0cm}
\caption{Three-dimensional visualization of a four-dimensional 
configuration in the deconfinement at 20 cooling steps  
with topological charge $Q=2$. 
The dots indicate the topological charge density $|q(x)| > 0.005$ and the
lines represent monopole loops.} 
\end{figure}
Similar phenomena were discussed by other groups 
on semiclassical configurations \cite{andere}.
This might indicate that both confinement mechanisms have a  common 
origin and that both approaches can be united.
It is believed that instantons and also monopoles can explain chiral
symmetry breaking \cite{SHU88,MIA95}. 
In this contribution we study the origin of the relation between 
the topological objects by analyzing the correlation functions 
per gauge-configuration and by visualizing the topological structure 
by means of 3D graphics.  

To investigate monopole currents we project $SU(N)$
onto its  abelian degrees of freedom, such that an abelian $U(1)^{N-1} $
theory remains \cite{thooft2}. We employ the
so-called maximum abelian gauge being most favorable for our
purposes.
For the definition of the monopole currents $m_{i}(x,\mu), i = 1,...,N,$ 
we use the standard method \cite{SCH87}. 
From the monopole currents we define the local monopole density as
$ \rho(x) = \frac{1}{ 4 N V_{4}} \sum_{\mu,i} | m_{i}(x,\mu) | $.  
%

There exist several definitions of the  topological charge on the lattice.
The field theoretic prescriptions are a straightforward  discretization of
the continuum expression.
To get rid of the renormalization 
constants we apply the ``Cabbibo-Marinari cooling method'' 
which smooths the quantum fluctuations of a gauge-field. 
Other  topological charge operators can be obtained from the 
geometric definitions. The discrete set of link 
variables is interpolated to the continuum and then the topological charge
is calculated directly.  
Concerning the correlation between monopoles 
and instantons it  was shown in \cite{letter96} that the 
geometric L\"uscher charge definition yields qualitatively the same 
results as the field theoretic prescriptions.  
Therefore we employ in these  studies of the topological charge density $q(x)$ 
the field theoretic plaquette and hypercube prescription \cite{divecchia}.
To measure correlations between topological quantities
we calculate functions of the type 
\begin{eqnarray} \label{correlations}
 q(0) q(r)      ,
 \rho(0) |q(r)|  
\end{eqnarray}
per gauge-field configuration.

Our simulations were performed on a $12^{3} \times 4$ lattice with
periodic boundary conditions using the Metropolis algorithm.
The observables were studied  both in the confinement
and the deconfinement phase of pure SU(2) theory at inverse gluon coupling
$\beta=4/g^{2}=2.25$ and $2.4$,  respectively.
 For each run  we made 100  measurements, separated by 100  iterations. 
In  previous work we found a spatial relation between instantons and  
monopoles  by averaging  the correlation between the topological 
charge density and the monopole density over an ensemble of gauge-fields 
\cite{letter96}. The coexistence 
between both topological objects turned out to be hardly affected by cooling 
and behaved similarly in both phases of the theory.  
In this contribution we study  the origin of the nontrivial correlation 
between monopoles and instantons by considering single gauge-fields. 
In particular we are interested in the 
reason for the similarity in both phases.
 
Fig.~1 presents the auto-correlations of the topological charge in the 
plaquette definition and the $\rho|q|$-correlations after 20 cooling 
steps for 100 independent configurations. 
In the confinement phase the auto-correlation functions have many different 
amplitudes reflecting a large variety of topologically nontrivial 
configurations. Also the 
corresponding monopole-instanton correlations show many different amplitudes. 
In the deconfinemet phase only about 15 \% of the auto-correlation functions 
are nontrivial. 
All of these configurations give rise to a nontrivial $\rho|q|$-correlation.
This indicates that the relation between monopoles and instantons found on 
gauge average also holds for single configurations.  

In Fig. 2 we  visualize this relationship 
by directly displaying clusters of topological charge and
by drawing monopole loops for fixed $t$ or $x$ of a specific configuration 
in the deconfinement phase.
For any value of the topological charge density $q(x) > 0.005$ a light dot
and for $q(x) < -0.005$ a dark dot is plotted.
Monopole loops are represented by lines.
This configuration has a total  topological 
charge of $Q=2$. One observes a timelike monopole 
loop passing through each instanton.  

To summarize, we 
analyzed the topological structure of the $SU(2)$ vacuum at finite 
temperature in both phases of the theory. 
The auto-correlation functions of the topological charge density are nontrivial 
after 20 cooling steps reflecting the existence of instantons. 
Correlation functions between the topological charge density 
and the 
monopole density are hardly affected by cooling indicating a close spatial
relation between instantons and monopoles. This observation also holds 
in the deconfinement phase. 

Studying the origin of the coexistence between 
monopoles and instantons in more detail we computed correlation 
functions per configuration. In the deconfinement phase only approximately
15 \% of the configurations carry a topological charge and all of 
these configurations give rise to nonvanishing $\rho|q|$-correlations.
Visualization demonstrated that across the transition and after  cooling 
instantons are squeezed in time direction and are accompanied by timelike 
monopoles. It will be interesting to check this result by other methods not 
relying on cooling like inverse blocking.


\vspace{-0.2cm}

\begin{thebibliography}{99}
\itemsep=0cm

\vspace{-0.1cm}

\bibitem{wir}
S.~Thurner, H.~Markum and W.~Sakuler, Proceedings of Confinement 95,
Osaka 1995, eds. H.~Toki et al. (World Scientific, 1996) 77 (hep-th/9506123);
S.~Thurner, M.~Feurstein, H.~Markum and W.~Sakuler,
Phys.~Rev.~{D54} (1996) 3457;
M.~Feurstein, H.~Markum and S.~Thurner, 
Nucl. Phys. B (Proc. Suppl.) 53 (1997) 553.
\bibitem{andere}
M.N. Chernodub and F.V. Gubarev, JETP Lett. 62 (1995) 100; 
A.~Hart and M.~Teper, Phys.~Lett.~{ B371} (1996) 261;
V.~Bornyakov and G.~Schierholz, Phys.~Lett.~{B384} (1996) 190;
R.C.~Brower, K.N.~Orginos and Chung-I~Tan, 
Phys.~Rev.~D55 (1997) 6313.
\bibitem{SHU88}
E.V.~Shuryak, Nucl.~Phys.~{B302} (1988) 559.
\bibitem{MIA95}
O. Miyamura, Nucl. Phys. {B} (Proc.~Suppl.) {42} (1995) 538.
\bibitem{thooft2}
G.~'t~Hooft, Nucl.~Phys.~{B190} (1981) 455.
\bibitem{SCH87}
A.S.~Kronfeld, G.~Schierholz and U.-J.~Wiese, 
Nucl.~Phys. {B293} (1987) 461.
\bibitem{letter96}
M.~Feurstein, H.~Markum and S.~Thurner,
Phys.~Lett.~{B396} (1997) 203.
\bibitem{divecchia}
P.~Di~Vecchia, K.~Fabricius, G.C.~Rossi and G.~Veneziano,
Nucl.~Phys. {B192} (1981) 392;
Phys.~Lett. {B108} (1982) 323;
Phys.~Lett. {B249} (1990) 490.
\end{thebibliography}
\end{document}